\documentclass[prl,twocolumn,preprintnumbers,amsmath,amssymb]{revtex4}
\usepackage{graphicx}
\usepackage{latexsym}
\usepackage{amsmath}
\usepackage{graphics}
\usepackage{amssymb}
\usepackage{layout}
\usepackage{verbatim}
\usepackage{amsfonts,epsfig}

\newcommand{\beq}{\begin{equation}}
\newcommand{\eeq}{\end{equation}}
\newcommand{\bea}{\begin{eqnarray}}
\newcommand{\eea}{\end{eqnarray}}
\newcommand{\nn}{\nonumber}
\newcommand{\tr}{\hbox{Tr}}

\begin{document}

\title{Analytically solvable driven time-dependent two-level quantum systems}
\author{Edwin Barnes}
\author{S. Das Sarma}
\affiliation{Condensed Matter Theory Center, Department of Physics, University of Maryland, College Park, MD 20742-4111, USA}

\begin{abstract}
Analytical solutions to the time-dependent Schr\"odinger equation describing a driven two-level system are invaluable to many areas of physics, but they are also extremely rare. Here, we present a simple algorithm that generates an unlimited number of exact analytical solutions. We show that a general single-axis driving term and its corresponding evolution operator are determined by a single real function which is constrained only by a certain inequality and initial conditions. Any function satisfying these constraints yields an exact analytical solution. We demonstrate this method by presenting several new exact solutions to the time-dependent Schr\"odinger equation. Our general method and many of the new solutions we present are particularly relevant to qubit control in quantum computing applications.
\end{abstract}

\maketitle
The search for analytically solvable driven two-level quantum systems began shortly after the birth of quantum mechanics and continues into the present. Such systems are ubiquitous throughout quantum physics, and it is notoriously difficult to acquire exact analytical solutions to the relevant time-dependent Schr\"odinger equation aside from a few special cases. Perhaps the most famous examples of exactly soluble two-level evolution are the Landau-Zener \cite{Landau_PZS32,Zener_PRSL32} and Rabi \cite{Rabi_PR37} problems. The latter of course has several generalizations including the paradigmatic Jaynes-Cummings model \cite{Jaynes_PIEEE63}. Another well known exact solution discovered in the early 1930s by Rosen and Zener is the hyperbolic secant pulse \cite{Rosen_PR32}. This solution has proven important in the contexts of self-induced transparency \cite{McCall_PR69} and qubit control \cite{Economou_PRB06,Greilich_NP09,Poem_PRL11}, and has since been extended to a family of analytical controls by a number of authors \cite{Bambini_PRA81,Bambini_PRA84,Hioe_PRA84,Zakrzewski_PRA85,Silver_PRA85,Robinson_PRA85,Ishkhanyan_JPhysA00,Kyoseva_PRA05,Vitanov_NJP07,Hioe_Wiley07}. Some additional classes of soluble pulses were also discovered recently \cite{Jha_PRA10,Gangopadhyay_PRB10}, as well as a few oscillatory control examples \cite{Gangopadhyay_PRB10,Xie_PRA10}.

Despite these isolated successes, analytically solvable two-state problems have remained extremely rare. This fact has become more poignant in recent decades with the advent of quantum computation, where analytically solvable control pulses are especially attractive in light of the many advantages they offer in relation to the design of qubit control operations. In particular, such solutions can facilitate the development of controls that are both precise and robust without the need for long control sequences \cite{Economou_PRB06,Greilich_NP09,Economou_PRB12,Motzoi_PRL09,Chow_PRA10,Gambetta_PRA11}. However, the relatively small number of known analytically-solvable control fields greatly limits the options one has when adopting an analytical approach to qubit gate design, and it is unlikely that any of the known examples will be ideal for a specific situation, especially given that almost all of the these predate the inception of quantum computing.

In this Letter, we present a completely new theoretical approach to the driven two-state problem. We derive an algorithm that produces an unbounded number of analytically solvable two-level systems driven by a single-axis control field. We develop this method by showing that a general single-axis control field and its associated evolution operator are both determined by a single real function $q(t)$, and we give the explicit functional dependencies on $q(t)$. We further derive an inequality and initial conditions which $q(t)$ must obey in order for the resulting evolution operator to be a proper solution of the Schr\"odinger equation. Any $q(t)$ which satisfies these constraints corresponds to an analytically solvable two-state problem. We demonstrate our method by deriving several new analytical solutions. We also determine how properties of $q(t)$ translate to the control field and evolution operator. This `reverse-engineering' approach is especially appropriate in the context of quantum control where one typically wishes to achieve a particular evolution by applying a control field whose basic features are restricted only by a few experimentally-imposed constraints.

The Hamiltonian we will consider has the general form
\beq
H={J(t)\over2}\sigma_z+{h\over2}\sigma_x,\label{ham}
\eeq
where $J(t)$ is the control (driving) field, $h$ is a constant, and $\sigma_z$, $\sigma_x$ are Pauli matrices. This Hamiltonian describes any two-level system which is driven along a single axis (denoted by $z$) \footnote{The methods of this paper apply equally well to a two-axis control with $H={J(t)\over2}\cos(\omega t)\sigma_z+{J(t)\over2}\sin(\omega t)\sigma_y+{h\over2}\sigma_x$ where $\omega$ is a constant. One simply takes $h\to h+\omega$ in the various results we obtain.}. In many contexts, $h$ can be interpreted as the energy splitting between the two levels \cite{Greilich_NP09,Poem_PRL11,Martinis_PRL05}, but in other contexts, e.g. singlet-triplet qubits \cite{Petta_Science05,Foletti_NP09,Maune_Nature12,Wang_arxiv12}, $J(t)$ could be thought of as a time-varying energy splitting between the states. We parametrize the evolution operator corresponding to $H$ as
\beq
U=\left(\begin{matrix}u_{11} & -u_{21}^*\cr u_{21} & u_{11}^*\end{matrix}\right), \qquad |u_{11}|^2+|u_{21}|^2=1,\label{defofU}
\eeq
and we transform to a rotating frame in the $x$-basis:
\beq
D_\pm={1\over\sqrt{2}}e^{\pm iht/2}(u_{11}\pm u_{21}).\label{Dumap}
\eeq
The functions $D_\pm$ then solve the following set of equations which follow from the Schr\"odinger equation for the evolution operator $U$:
\beq
\dot D_\pm=-i{J\over2}e^{\pm iht}D_\mp.\label{Dschrod}
\eeq
These equations can be combined to yield a second-order differential equation for $D_+$:
\beq
\ddot D_++(-ih-\dot J/J)\dot D_++(J^2/4)D_+=0.\label{Dpteqn}
\eeq
At this point, one typically inserts a particular expression for $J(t)$ and then attempts to solve this equation for $D_+(t)$ to obtain the corresponding evolution operator. Indeed, this is the manner in which most of the previously known analytical solutions were found. In some cases, the precise form of $J(t)$ was dictated by the physics of the problem, in other cases $J(t)$ was chosen so that Eq. (\ref{Dpteqn}) became a well known differential equation \footnote{In a few cases, integrability conditions led to a specific differential equation for the control field. See e.g. \cite{Gangopadhyay_PRB10}.}.

We adopt a dramatically different approach which begins by noticing that we can also view Eq. (\ref{Dpteqn}) as a differential equation for $J(t)$. It turns out that this equation can be solved exactly for arbitrary $D_+$:
\beq
J(t)=\pm{\dot D_+ e^{-iht}\over\sqrt{c-{1\over4}D_+^2e^{-2iht}-{ih\over2}\int_0^t dt'e^{-2iht'}D_+^2(t')}},\label{JfromDp}
\eeq
where $c$ is an integration constant. Given this expression for $J(t)$, Eq. (\ref{Dschrod}) then gives $D_-$ in terms of $D_+$:
\beq
D_-=\pm 2i \sqrt{c-{1\over4}D_+^2e^{-2iht}-{ih\over2}\int_0^t dt'e^{-2iht'}D_+^2(t')}.\label{DmfromDp}
\eeq
Supposing the evolution begins at $t=0$, we impose $D_+(0)=D_-(0)=1/\sqrt{2}$, which in turn implies that $c=0$ and that we should choose the minus sign in (\ref{DmfromDp}). Note that imposing the initial condition at $t=0$ will not prevent us from obtaining solutions that span any range of the time domain as we will clarify later on.

Our results so far can be interpreted as a reverse-engineering of the control $J(t)$: we can choose the evolution by picking the function $D_+$ as we like, and then use (\ref{JfromDp}) and (\ref{DmfromDp}) to determine $D_-(t)$ and $J(t)$. However, we must ensure that unitarity is preserved: $|D_+|^2+|D_-|^2=1$. This is automatically satisfied by the general ansatz
\bea
D_+&=& e^{i(F-K+ht)}\cos\Phi,\nn\\
D_-&=& e^{-iK}\sin\Phi,
\eea
where $F$, $K$, and $\Phi$ are arbitrary real functions. These expressions lead to the following forms for $u_{11}$ and $u_{21}$:
\bea
u_{11}&=&{1\over\sqrt{2}}e^{i(ht/2-K)}(e^{iF}\cos\Phi+\sin\Phi),\nn\\
u_{21}&=&{1\over\sqrt{2}}e^{i(ht/2-K)}(e^{iF}\cos\Phi-\sin\Phi).\label{u11u21}
\eea
The initial conditions on $D_+$, $D_-$ translate to $\Phi(0)=\pi/4$, $F(0)=K(0)=0$. Eq. (\ref{DmfromDp}) imposes relations between $F$, $K$, and $\Phi$ which can be extracted by first squaring both sides of this equation, differentiating the result with respect to time, and then equating the real and imaginary parts of both sides to arrive at
\bea
\dot F+h&=&\dot K(1-\tan^2\Phi),\nn\\
\dot \Phi&=&\dot K\tan F\tan\Phi.\label{nonlineqns}
\eea
In terms of these functions, $J(t)$ can be expressed as
\beq
J(t)=2\dot K\sec F\tan\Phi.\label{JfunofKFPhi}
\eeq
These relations further fix some additional initial conditions: $\dot\Phi(0)=0$, $\dot F(0)=-h$, and $J(0)=2\dot K(0)$. $\dot K(0)$ is not restricted by these relations. The next step is to notice that we can solve (\ref{nonlineqns}) explicitly for $\Phi$ and $K$ in terms of $F$. For $\Phi$ we obtain
\beq
\sin(2\Phi)=\sec F e^{h\int_0^t dt'\tan F(t')},\label{PhiFrel}
\eeq
where we have already chosen the integration constant so that the initial conditions are satisfied. One may then use either of the two equations in (\ref{nonlineqns}) to solve for $K$. It then follows that once the function $F(t)$ is specified, so are the control field and its evolution operator.

As a first check we may consider the case $h=0$. In this case, the only solution to Eq. (\ref{PhiFrel}) is to choose $\Phi=\pi/4$ and $F=0$. $\dot K$ is then unconstrained by (\ref{nonlineqns}), and from Eq. (\ref{u11u21}), it is clear that we obtain a $z$-rotation for any $K={1\over2}\int_0^tdt'J(t')$. We may also consider the case $J=0$, which is realized by setting $\dot K=0$, implying that $K=0$. It follows immediately from (\ref{nonlineqns}) that $\dot\Phi=0$ and $\dot F=-h$, so that $\Phi=\pi/4$ and $F=-ht$. This solution is of course consistent with Eq. (\ref{PhiFrel}), and plugging into (\ref{u11u21}) reveals a free precession about the $x$-axis as expected.

We stress that Eq. (\ref{PhiFrel}) is not simply an equation which gives $\Phi$ once $F$ is chosen. This equation actually places strong constraints on $F$ as can be seen by noticing that the RHS does not generically respect the upper and lower bounds on $\sin(2\Phi)$; generic choices of $F$ will yield a RHS which exceeds unity. This restriction on $F$ reflects the fact that our Hamiltonian, Eq.~(\ref{ham}), generates only a subset of all possible trajectories on the Bloch sphere.

It is helpful to replace $F$ by a new function $q$:
\beq
F=\arctan\left({\dot q\over hq}\right),\label{Ffromq}
\eeq
in which case one finds
\beq
\sin(2\Phi)=\sqrt{q^2+\dot q^2/h^2},\label{Phifromq}
\eeq
\beq
\dot K={1\over2}{hq(\ddot q+h^2q)\over h^2q^2+\dot q^2}\left[1+{h\over\sqrt{h^2(1-q^2)-\dot q^2}}\right],\label{Kfromq}
\eeq
\beq
J={\ddot q+h^2q\over\sqrt{h^2(1-q^2)-\dot q^2}}.\label{Jfromq}
\eeq
The initial conditions on $F$, $K$ and $\Phi$ translate to
\beq
q(0)=1,\quad \dot q(0)=0,\quad \ddot q(0)=-h^2,\label{initconditions}
\eeq
and the requirement that the RHS of Eq. (\ref{Phifromq}) does not have a magnitude exceeding unity leads to
\beq
\dot q^2\le h^2(1-q^2).\label{inequality}
\eeq
Eqs.~(\ref{Ffromq})-(\ref{inequality}) together comprise the main result of this paper. Any function $q(t)$ which satisfies Eqs.~(\ref{initconditions}) and (\ref{inequality}) will produce an analytical solution to the Schr\"odinger equation, with the control field and its corresponding evolution operator given by Eqs.~ (\ref{Ffromq})-(\ref{Jfromq}). This simple prescription will enable us to generate an unlimited number of analytically solvable two-state problems along with their explicit solutions. We will now demonstrate this by writing down several new examples.

\begin{figure}
\begin{center}
\includegraphics[height=3cm, width=1\columnwidth]{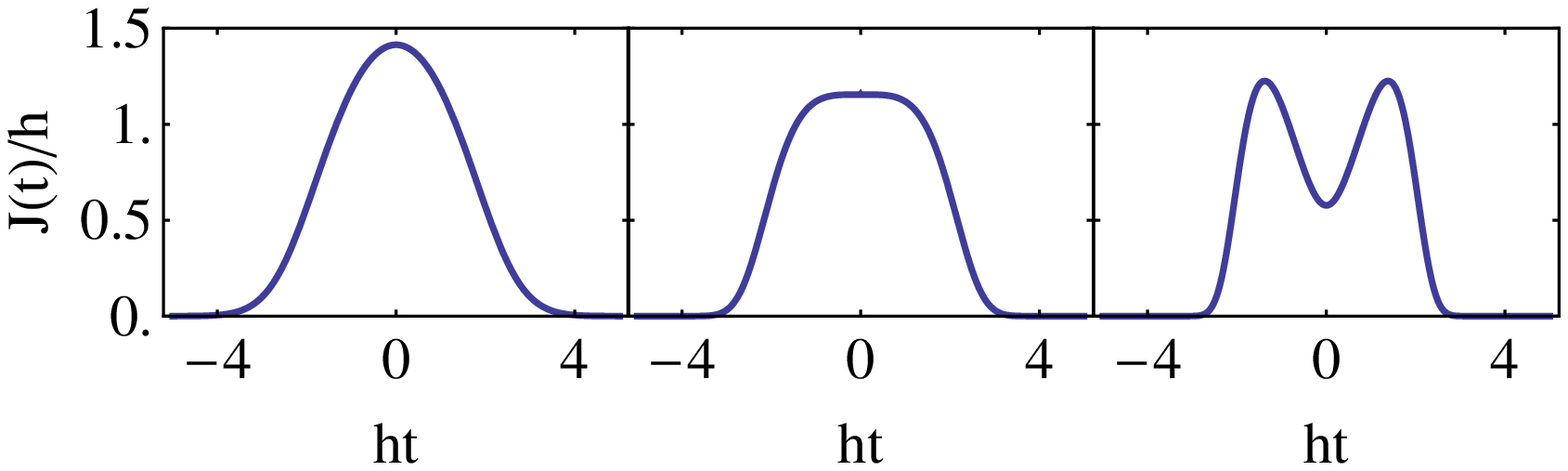}
\includegraphics[height=3cm, width=1\columnwidth]{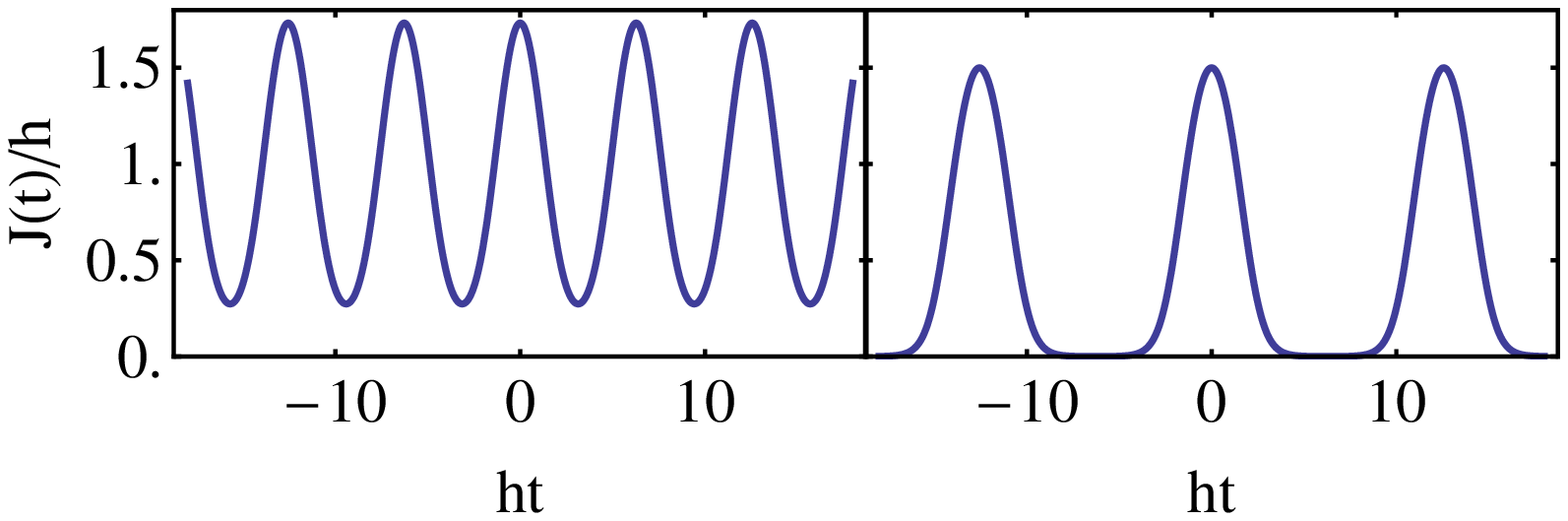}
\caption{\label{fig:example1} (Color online) Pulse from Eq. (\ref{example1J}) with (top panel from left to right) $a=0,{2\over3},{5\over3}$ and (bottom panel from left to right) $a=-1,-{1\over4}$.}
\end{center}
\end{figure}
There is a special choice of $q$ for which the inequality in Eq.~(\ref{inequality}) is saturated, namely $q=\cos(ht)$; this choice corresponds to a pure $x$-rotation with $J=0$. As a first, more nontrivial example, we consider the choice
\beq
q(t)=\exp\left\{-(2/a)\sinh^2(\sqrt{a}ht/2)\right\},\label{example1q}
\eeq
where $a$ is any real number satisfying $a\le2$. It can be verified that (\ref{example1q}) satisfies Eqs.~(\ref{initconditions}) and (\ref{inequality}). Eq.~(\ref{Jfromq}) then gives
\beq
J(t)={h\left[(1/a)\sinh^2(\sqrt{a}ht)-2\sinh^2(\sqrt{a}ht/2)\right]\over \sqrt{e^{(4/a)\sinh^2(\sqrt{a}ht/2)}-(1/a)\sinh^2(\sqrt{a}ht)-1}},\label{example1J}
\eeq
and this result is plotted for different values of $a$ in Fig.~\ref{fig:example1}. The corresponding evolution operator can be computed straightforwardly from Eqs.~(\ref{Ffromq})-(\ref{Kfromq}). An interesting feature of this solution is that it describes a single pulse when $a\ge0$ and an oscillatory control field when $a<0$. This solution illustrates that the complexity of $J(t)$ and $U(t)$ tends to be comparable to that of $q(t)$, in contrast to most of the analytical solutions known prior to this work, where simple pulse shapes often yield evolutions governed by special functions.

\begin{figure}
\begin{center}
\includegraphics[height=6cm, width=\columnwidth]{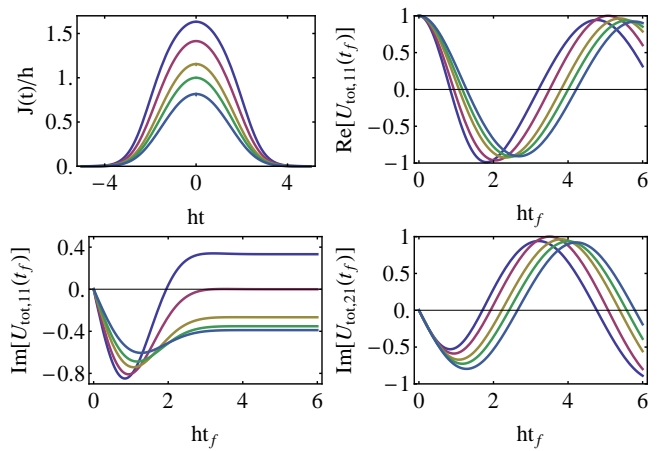}
\caption{\label{fig:gaussExample} (Color online) Pulse from Eq.~(\ref{gaussExampleJ}) with (from top to bottom) $b=-{1\over4},0,{1\over2},1,2$ and the corresponding evolution operator components.}
\end{center}
\end{figure}
A second new nontrivial example arises from the choice
\beq
q(t)={1\over1+b}[e^{-h^2t^2/2}+b\cos(ht)],\label{gaussExampleq}
\eeq
which upon inserting into Eq. (\ref{Jfromq}) yields
\beq
J(t)={h^3t^2e^{-h^2t^2/2}\over\sqrt{1-(1+h^2t^2)e^{-h^2t^2}+2b\chi(t)}},\label{gaussExampleJ}
\eeq
with $\chi(t)\equiv1-e^{-h^2t^2/2}[\cos(ht)+ht\sin(ht)]$. Plots of this pulse and its corresponding evolution operator for different values of $b$ are shown in Fig.~\ref{fig:gaussExample} where it is apparent that these are smooth, Gaussian-like pulses with $b$ controlling the magnitude and width. These pulses offer an attractive alternative to the widely used hyperbolic secant pulses \cite{Rosen_PR32}: they converge to zero much more rapidly and are thus more localized, and the corresponding evolution operator is expressed in terms of elementary functions rather than the hypergeometric function. Gaussian pulses have proven useful in the context of error suppression in superconducting qubit gate design \cite{Motzoi_PRL09,Chow_PRA10,Gambetta_PRA11}, however these were not analytically solvable and required numerical optimization techniques.

The two examples we have given so far already illustrate that properties of $q(t)$ are reflected in $J(t)$. In the context of quantum control, experimental constraints often require $J(t)$ to be smooth, bounded, and well localized in time. Smoothness of $J(t)$ is guaranteed by choosing a smooth $q(t)$. In addition, it is clear from Eq.~(\ref{Jfromq}) that a given $q(t)$ will produce a well defined pulse (i.e. $J(t)\to0$ as $t\to\pm\infty$) if $q\to0$ and $\ddot q\to0$ as $t\to\pm\infty$, or if $\ddot q\to -h^2q$ in this limit (the former condition is satisfied by Eq.~(\ref{example1q}) and the latter by Eq.~(\ref{gaussExampleq})). The strict inequality $0<\dot q^2< h^2(1-q^2)$ for $t>0$ further ensures that $J(t)$ will be a bounded function. In some cases (e.g. singlet-triplet qubits \cite{Wang_arxiv12}), one also needs to impose positive control, $J(t)\ge0$; as seen from Eq.~(\ref{Jfromq}), this requires $\ddot q\ge-h^2q$. Eq.~(\ref{example1q}) with $a\ge0$ and Eq.~(\ref{gaussExampleq}) satisfy the above criteria, in which case we have a single, positive, bounded pulse as confirmed by the top panels of Fig.~\ref{fig:example1} and the top left panel of Fig.~\ref{fig:gaussExample}. When $a<0$ in Eq.~(\ref{example1q}), $\dot q(t_0)=0$ for some $t_0>0$, leading to the periodic behavior shown in the lower panels of Fig.~\ref{fig:example1}.

The examples of Eqs.~(\ref{example1q}) and (\ref{gaussExampleq}) also exhibit the connection between $q(t)$ and the corresponding evolution operator. For instance, these examples contain even control functions, $J(-t)=J(t)$, which follows directly from the evenness of the chosen $q(t)$. For such pulses, the evolution operator given by Eqs.~(\ref{defofU}) and (\ref{u11u21}) describes half the evolution due to the pulse since we have imposed $U(0)=1$. If we wish to instead impose $U(-t_f)=1$ for some $t_f>0$, then the full evolution operator describing the evolution from $t=-t_f$ onward is $U_{tot}(t)=U(t)U^\dagger(-t_f)$. At $t=t_f$, one then finds that $\tr\{U_{tot}(t_f)\sigma_y\}=0$ and $\tr\{U_{tot}(t_f)\sigma_z\}\sim\sin(2\Phi(t_f))=\sqrt{q^2(t_f)+\dot q^2(t_f)/h^2}$, implying that the pulse effects a rotation about an axis in the $x-z$ plane which depends on the behavior of $q$ at $t=t_f$. For well defined pulses in particular, if $t_f$ is sufficiently large and $q\to A\cos(ht)+B\sin(ht)$ as $t\to\infty$, then the rotation axis is determined by $A$ and $B$, where $A=B=0$ yields an $x$-rotation. It is then clear from Eq.~(\ref{example1q}) that the pulse family of Eq. (\ref{example1J}) implements $x$-rotations for any $a\ge0$, while Eq.~(\ref{gaussExampleq}) reveals that the pulses given in Eq. (\ref{gaussExampleJ}) implement rotations about various axes in the $x-z$ plane depending on $b$. In particular, denoting the rotation axis by $(n_x,0,n_z)$ and the angle by $\theta$, the lower left panel of Fig.~\ref{fig:gaussExample} reveals that $\hbox{Im}[U_{tot,11}(t_f)]=-n_z\sin(\theta/2)$ quickly saturates to a constant $b$-dependent value beyond $ht_f\gtrsim 3$. Given $n_z$ and $\theta$, one can first choose $b$ to fix the combination $n_z\sin(\theta/2)$ and then tune $t_f>3/h$ to achieve the target rotation.

We can systematically find $q$'s which satisfy Eq. (\ref{inequality}) by first choosing a function ${\cal P}(q)$ such that $0\le{\cal P}(q)\le 1-q^2$ and then solving $\dot q^2=h^2{\cal P}(q)$. Since this equation is homogeneous in $q$, it can be integrated directly:
\beq
ht=\int_q^1 {dq'\over\sqrt{{\cal P}(q')}}\equiv {\cal W}(q).
\eeq
The fact that ${\cal P}(q)$ is strictly nonnegative guarantees that
\beq
{\cal W}(q)\ge \int_q^1{dq'\over\sqrt{1-q'^2}}=\arccos q.\label{Winequality}
\eeq
The initial conditions on $q(t)$ become the condition ${\cal W}(q)\to\sqrt{2-2q}$ as $q\to1$. Any invertible function ${\cal W}(q)$ satisfying this boundary condition and the inequality (\ref{Winequality}) automatically produces an analytical solution to the Schr\"odinger equation with $q(t)={\cal W}^{-1}(ht)$.

\begin{figure}
\begin{center}
\includegraphics[height=3cm, width=\columnwidth]{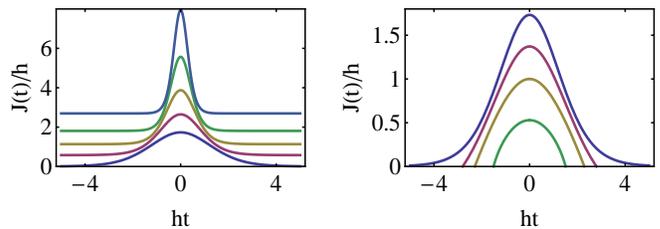}
\caption{\label{fig:example2} (Color online) Pulse from Eq.~(\ref{example2J}) with (left panel, from top to bottom) $a=2\sqrt{2},2,\sqrt{2},1,1/\sqrt{2}$ and (right panel, from top to bottom) $a=1/\sqrt{2},0.6,0.5,0.4,0.3$.}
\end{center}
\end{figure}
\begin{figure}
\begin{center}
\includegraphics[height=3cm, width=\columnwidth]{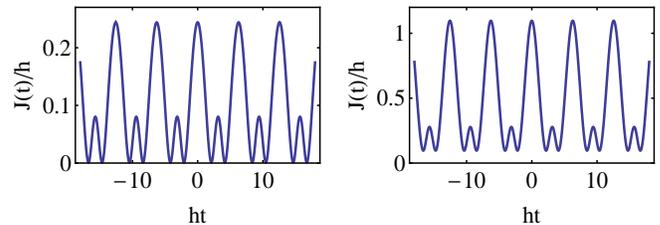}
\caption{\label{fig:example3} (Color online) Control from Eq.~(\ref{example3q}) with (left) $a=0.1$ and (right) $a=0.5$. }
\end{center}
\end{figure}
To give an example using this approach, we choose
\beq
{\cal W}(q)=(1/a)\hbox{arctanh}(a\sqrt{2-2q}),\label{example2W}
\eeq
where $a$ is a real constant. Upon inverting, we obtain
\beq
q(t)=1-{1\over 2a^2}\tanh^2(aht),\label{example2q}
\eeq
\beq
J(t)={h\left[14a^2-1+(2a^2-1)\cosh(2aht)\right]\hbox{sech}^2(aht)\over 2\coth(aht)\sqrt{4a^2\left[1-\hbox{sech}^4(aht)\right]-\tanh^2(aht)}}.\label{example2J}
\eeq
The corresponding control pulses are shown in Fig.~\ref{fig:example2}. The pulse with $a=1/\sqrt{2}$ asymptotes to zero at large times, while pulses with $a>1/\sqrt{2}$ asymptote to the positive constant $(2a^2-1)/\sqrt{4a^2-1}$. These latter pulses could be relevant in experimental situations in which it is not possible to turn off $J(t)$ completely, as in the case of singlet-triplet qubits \cite{Petta_Science05,Foletti_NP09,Maune_Nature12}. Pulses with $a<1/\sqrt{2}$ are defined on a strictly finite time interval as shown in the right panel of Fig.~\ref{fig:example2}.

As a final example, we make the choice
\beq
{\cal W}(q)=\cos^{-1}\left\{1-\left(a+1/a\right)\left[\tan^{-1}(aq)-\tan^{-1}a\right]\right\},\nonumber
\eeq
yielding
\beq
q(t)=(1/a)\tan\left\{\tan^{-1}a-(2a/(1+a^2))\sin^2(ht/2)\right\}.\label{example3q}
\eeq
The resulting $J(t)$ is a straightforward combination of elementary functions that is plotted in Fig.~\ref{fig:example3}, where it is apparent that this control resembles two superimposed sinusoids. Varying the parameter $a$ changes the amplitude and average value of these sinusoids, but not their frequency. This type of solution is relevant, e.g., for determining the properties of two-level fluctuators which cause decoherence in superconducting qubits \cite{Martinis_PRL05,Gangopadhyay_PRB10}.

In conclusion, we have shown how to systematically obtain an unlimited number of analytically solvable controls and have provided explicit analytical formulas for their corresponding evolution operators. This vastly increases the number of known analytical solutions to the two-state Schr\"odinger equation, providing a powerful tool in robust quantum gate design and in the myriad of other physical contexts in which the two-state problem arises.

This work is supported by LPS-NSA and IARPA.

\end{document}